\documentclass{article}
\usepackage{amssymb, amsmath, latexsym, amsfonts}
\usepackage[dvips]{epsfig}
\newtheorem{proposition}{Proposition}
\newtheorem{corollary}{Corollary}
\newtheorem{definition}{Definition}
\newtheorem{notation}{Notation}
\newtheorem{theorem}{Theorem}
\begin{document}
\title{\bf Collision-resistant hash function based on composition of functions}
\author{Ren\'e Ndoundam${}^{a, b}$, Juvet karnel Sadi\'e${}^{a,b}$   \\
${}^{a}${\small Universit\'e de Yaound\'e I, UMI 209, UMMISCO, B.P. 337 Yaound\'e, Cameroun}  \\
${}^{b}${\small Universit\'e de Yaound\'e I, LIRIMA, Equipe GRIMCAPE, D\'epartement d'Informatique, B.P. 812 Yaound\'e} \\
{\small E.mail : ndoundam@gmail.com, ndoundam@yahoo.com , karnel12@yahoo.fr  } \\}

\maketitle
\begin{abstract}
A cryptographic hash function is a deterministic procedure that compresses an arbitrary block of numerical data and returns a fixed-size bit string. There exists many
hash functions: MD5, HAVAL, SHA, ... It was reported that these hash functions are no longer secure.  Our work is focused on the construction of a new hash function 
 based on composition of functions. The construction used the NP-completeness  of Three-dimensional contingency tables and the relaxation of the  constraint that a 
 hash function should also be a compression function.\\
\end{abstract}

{\bf Keywords :}
NP-complete, One-way function, Matrix of zeros and ones, Three-dimensional contingency table, Collision-resistant hash function.

\pagestyle{myheadings}
\thispagestyle{plain}

\section{Introduction}
A cryptographic hash function is a deterministic procedure that compresses an arbitrary block of data and returns fixed-size bit string,
 the hash value (message digest or digest). An accidental or intentional change of the data will almost certainly change the 
 hash value. Hash functions are used to  verify  the integrity of data or data signature. \\
Let us suppose that $h : X \rightarrow Y$ is a hash function without key. The function $h$ is secured if the following three
 problems are difficult to solve. \\

{\bf Problem 1:} First Preimage attack \\
$\phantom{salutsa}$ {\it Instance:} a function $h : X \rightarrow Y$ and an image  $y  \in Y$   \\
$\phantom{salutsa}$ {\it Query:} $\phantom{sa}$ $x \in X$ such that $h(x)=y$   \\

We suppose that a possible hash $y$ is given, we want to know if there exists $x$ such that $h(x)=y$. If we can solve
 {\it First Preimage attack}, then $(x, y)$ is a valid pair. A hash function for which {\it First Preimage attack} can't be solved
 efficiently is sometimes called {\it Preimage resistant}. \\ 

{\bf Problem 2:}  Second Preimage attack \\
$\phantom{salutsa}$ {\it Instance:} a function  $h : X \rightarrow Y$ and an element  $x_1  \in X$      \\
$\phantom{salutsa}$ {\it Query:} $\phantom{sa}$ $x_2 \in X$ such that $x_1 \ne x_2$  and $h(x_1) = h(x_2)$   \\

A message $x_1$ is given, we want to find a message $x_2$ such that $x_2 \ne x_1$ and $h(x_1) = h(x_2)$. If this is possible, then $(x_2, h(x_1))$ is a valid pair. A function for which {\it Second preimage attack} can't be solve efficiently is sometimes called 
 {\it Second preimage resistant}.  \\

{\bf Problem 3:}  Collision attack\\
$\phantom{salutsa}$ {\it Instance:}     a function  $h : X \rightarrow Y$   \\
$\phantom{salutsa}$ {\it Query:} $\phantom{sa}$ $x_1 , x_2 \in X$ such that $x_1 \ne x_2$ and $h(x_1) = h(x_2)$  \\
We want to known if it is possible to find two distinct messages $x_1$ and $x_2$ such that $h(x_1) = h(x_2)$. A function for which Collision attack can't be solve efficiently is sometimes called {\it Collision resistant}.  \\

There exists many hash functions: MD4, MD5, SHA-0, SHA-1, RIPEMD, HAVAL. It was reported that such widely hash functions are no
 longer secured \cite{HYXW 07, HY 05, XHY 05, XW 05, XXW 05, XWW 05, BA 94}. Thus, new hash functions should be studied. The existing 
 hash functions such as MD4, MD5, SHA-0, SHA-1, RIPEMD, HAVAL... want to achieve two goals at the same time:
\begin{description}
\item[$a^\circ$)]   For any input $x$, they return a hash of $x$ (of fixed length, this length depends on the hash function choosed)
\item[$b^\circ$)]   Preimage resistant, Second Preimage resistant and Collision Resistant.
\end{description}
 
Our contribution is to separate the two goals defined in points $a^\circ$) and $b^\circ$). Our hash function $H_3$ is defined as follows:
\begin{itemize}
 \item  $H_3 \ = \ H_2 o H_1$,
 \item  $H_2$ is a classical hash function such as MD5, SHA-0, SHA-1, RIPEMD, HAVAL, ....
 \item  Given a  $y$,  find  $x$  such  that  $H_1(x) = y$  is  NP-Complete,
 \item  Find  $x_1$, $x_2$  such  that  $x_1 \ne  x_2$ and $H_1(x_1) =  H_1(x_2)$  is  NP-Complete,
 \item For any input $x$, the length of $H_1(x)$ is not fixed. This is the main difference with the classical hash functions.
\end{itemize}

The paper is organized as follows: in Section 2, some  preliminaries  are presented.  Section 3 is devoted to the  design  of our hash function. 
 Concluding remarks are stated  in Section 4. 
\section{Preliminaries}

Let's define some preliminaries useful for the next section.
 
\subsection{Two-dimensional}

Data security in two dimension have been studied by many authors \cite{Cox 80, Fel 72, Kao 93,San 84}.  Let $m$ and $n$ be two positive integers, and let $R=(r_1, r_2, \dots, r_m)$ and 
$S=(s_1, s_2, \dots, s_n)$ be non-negative integral vectors. Denoted by $\mathfrak{A}(R, S)$ the set of all $m \times n$ matrices $A=(a_{ij})$ satisfying
\begin{equation}  \notag
a_{ij} = 0 \ \ {\rm or} \ \ 1 \ \ {\rm for} \ \ i=1,2, \dots, m \ \ {\rm and} \ \ j=1,2, \dots, n ; 
\end{equation}
\begin{align} 
 \sum_{j=1}^{n}  a_{ij}  &= r_i \ \ {\rm for} \ \ i = 1, 2, \dots, m ;  \notag  \\
 \sum_{i=1}^{m}  a_{ij}  &= s_j \ \ {\rm for} \ \ j = 1, 2, \dots, n.    \notag 
\end{align}

Thus a matrix of 0's and 1's belongs to $\mathfrak{A}(R, S)$ provided its row sum vector is $R$ and its column sum vector is $S$. The
set $\mathfrak{A}(R, S)$ was studied by many authors \cite{Bru 80, Ful 60, Gal 57, Hab 63, Rys 57}. Ryser \cite{Rys 57} has defined an
 {\it interchange} to be a transformation which replaces the $ 2 \times 2$ submatrix : 
\begin{equation*}      
  B_0 =  \left(      
      \begin{array} {cc}
       1  &  0    \\   
       0  &  1
  \end{array}  
  \right) 
\end{equation*}   
of a matrix A of 0's and 1's with the  $2 \times 2$ submatrix    
\begin{equation*}    
  B_1 = \left(      
      \begin{array} {cc}    
       0  &  1    \\   
       1  &  0
  \end{array}  
  \right) 
\end{equation*}         
If the submatrix $B_0$ (or $B_1$) lies in rows $k, l$  and columns $u, v$, then we call the interchange a
 $(k, l ; u, v)$ {\it -interchange}. An interchange (or any finite sequence of interchanges) does not alter the row and column sum vectors
 of a matrix. Ryser has shown the following result.
\begin{theorem} \label{th:n1} \cite{Rys 57}
Let $A$ and $A^*$ be two $m$ and $n$ matrices composed of 0's and 1's, possessing equal row sum vectors and equal column sum vectors.
Then $A$ is transformable into $A^*$ by a finite number of interchanges.
\end{theorem}

Let us consider a matrix $A \in \{ 0 , 1 \}^{n \times n} \in \mathfrak{A}(R, S)$, i.e. its row sum vector $R$ is such that
 $R \in \{0, 1, 2, \dots, n \}^n$ and its column sum vector $S$ is such that $S \in \{0, 1, 2, \dots, n \}^n$. We define the function
 $g_1$ from $\{ 0 , 1 \}^{ n \times n}$ to $\mathbb{N}^{2n}$ as follows:
\begin{align}
g_1(A) =  &R(1)  ||  R(2) || \dots ||  R(n) ||   \notag  \\
                 &S(1)  || S(2)  || \dots || S(n)              \notag 
\end{align}
where $\|$ denotes the concatenation.

\subsection{Three-dimensional}

Irving and Jerrum \cite{IJ 94} have studied the extension of the problem in three dimension and shown that problems that are solvable in polynomial time in the two-dimensional case become NP-Complete. 
Suppose that for a given $n \times n \times n$ table $D$ of non-negative integers, and for each $i, j, k$, the row, column and file sums are denoted by $R(i,k), C(j, k)$ and $F(i,j)$ respectively. In other words:
\begin{align}
R(i,k)  &= \sum_{j=1}^{n}  D(i,j,k)   \notag   \\
C(j,k)  &= \sum_{i=1}^{n}  D(i,j,k)   \notag   \\
F(i,j)  &= \sum_{k=1}^{n}  D(i,j,k)   \notag 
\end{align}

The following problem is studied by Irving and Jerrum \cite{IJ 94} :   \\

{\bf Problem 4.}  Three-dimensional contingency tables (3DCT)   \\
$\phantom{salutsa}$ {\it Instance:}  A positive integer $n$, and for each $i, j, k$ non-negative integers   \\
$\phantom{salutsa Instancejgjkgj:}$ values $R(i,k)$, $C(j,k)$  and $F(i,j)$  \\
$\phantom{salutsa}$ {\it Question:}  Does there exist an $n \times n \times n$  contingency table $X$ of non-negative integers such that:
\begin{align}
  \sum_{j=1}^{n}  X(i,j,k)  &= R(i,k)    \notag       \\
  \sum_{i=1}^{n}  X(i,j,k)  &= C(j,k)    \notag       \\
  \sum_{k=1}^{n}  X(i,j,k)  &= F(i,j)    \notag   
\end{align}
for all $i, j, k$ ?  Irving and Jerrum show the following result:

\begin{corollary}  \label{corrol:1}  \cite{IJ 94}
3DCT is NP-Complete, even in the special case where all the row, column and file sums are 0 or 1.
\end{corollary}

Let us consider a matrix $A \in \mathbb{N}^{n \times n \times n}$ such that its row sum matrix is a matrix $R$ such that
 $R \in \mathbb{N}^{n \times n}$ (i.e. $R(i, k) \in \mathbb{N}$)
, the column sum matrix is a matrix $C$ such that $C \in \mathbb{N}^{n \times n}$ (i.e. $C(j, k) \in \mathbb{N}$) and
 the file sum matrix is a matrix $F$ such that $F \in \mathbb{N}^{n \times n}$ (i.e. $F(i, j) \in \mathbb{N}$).
 We define the function $g_2$ as follows:
\begin{equation}
g_2 \ : \ \mathbb{N}^{ n \times n \times n }  \longrightarrow  \mathbb{N}^{3n^2}   \notag
\end{equation}
\begin{align}
g_2(A) =  &R(1,1)  ||  R(1,2)   || \dots || R(n,n)   ||   \notag  \\
                 &C(1,1)   || C(1,2)   || \dots || C(n,n)   ||   \notag  \\       
                 &F(1,1)   ||  F(1,2)   || \dots || F(n,n)         \notag 
\end{align}

Let us consider the following matrices $A$ and $B$. We define the element product of matrices $A$ and $B$ as follows:

\begin{definition}  {\it Element Product of Matrices of dimension 2}   \\
Let $A, B \in \mathbb{R}^{n_1 \times n_2}$, we define the {\it Element product of matrices} $A$ and $B$ as follows:
\begin{multline}
C = \ A \ .* \ B  \ ; \  {\rm where} \ c_{ij} = a_{ij} \times b_{ij} \ {\rm for \ i, j \ such \ that}  \\ 1 \leq i \leq n_1 \ {\rm and}
                                                           \ 1 \leq j \leq n_2     \notag
\end{multline}
\end{definition}

\begin{definition} {\it Element Product of Matrices of dimension 3}   \\
Let $A, B \in \mathbb{R}^{n_1 \times n_2 \times n_3}$, we define the {\it Element product of matrices} $A$ and $B$ as follows:
\begin{multline}
C = \ A \ .* \ B \ ; \ {\rm where} \ c_{ijk} = a_{ijk} \times b_{ijk} \ {\rm for \ i,j,k \ such \ that}  \\
               1 \leq i \leq n_1 \ , \ 1 \leq j \leq n_2 \ {\rm and} \ 1 \leq k \leq n_3   \notag
\end{multline}
\end{definition}
  
\section{Design of the hash function}
Before the construction of our hash function, let us explain the main idea. 

\subsection{Explanation of the idea by an example}

In page 175 of paper \cite{Bru 80}, Brualdi gives the example of the following five matrices:

\begin{equation*}
 A_1 =  \left(      
      \begin{array} {ccc}
       1  &  1  & 0   \\   
       1  &  1  & 0   \\
       0  &  0  & 1
  \end{array}  
  \right) \ ; \
 A_2 =  \left(      
      \begin{array} {ccc}
       1  &  1  & 0   \\   
       1  &  0  & 1   \\
       0  &  1  & 0
  \end{array}  
  \right)  \ ; \
 A_3 =  \left(      
      \begin{array} {ccc}
       1  &  1  & 0   \\   
       0  &  1  & 1   \\
       1  &  0  & 0
  \end{array}  
  \right) 
\end{equation*} 
\begin{equation*}
 A_4 =  \left(      
      \begin{array} {ccc}
       0  &  1  & 1   \\   
       1  &  1  & 0   \\
       1  &  0  & 0
  \end{array}  
  \right) \ ; \
 A_5 =  \left(      
      \begin{array} {ccc}
       1  &  0  & 1   \\   
       1  &  1  & 0   \\
       0  &  1  & 0
  \end{array}  
  \right)  
\end{equation*} 
which belong to $\mathfrak{A}(R, S)$ where $R=S=(2,2,1)$. Let us note $W$ the following  matrix:
\begin{equation*}
 W =  \left(      
      \begin{array} {ccc}
       1  &  4  & 9   \\   
       2  &  8  & 18   \\
       3  &  12  & 27
  \end{array}  
  \right)  
\end{equation*} 
Based on the {\it Element Product of Matrix} defined in the previous subsection, it is easy to verify that:
\begin{equation*}
 A_1 \ .* \ W =  \left(      
      \begin{array} {ccc}
       1  &  4  & 0   \\   
       2  &  8  & 0   \\
       0  &  0  & 27
  \end{array}  
  \right) \ ; \
 A_2 \ .* \ W =  \left(      
      \begin{array} {ccc}
       1  &  4  & 0   \\   
       2  &  0  & 18   \\
       0  &  12  & 0
  \end{array}  
  \right)  \ ; \
 A_3 \ .* \ W =  \left(      
      \begin{array} {ccc}
       1  &  4  & 0   \\   
       0  &  8  & 18   \\
       3  &  0  & 0
  \end{array}  
  \right) 
\end{equation*}

\begin{equation*}
A_4 \ .* \ W = \left(      
      \begin{array} {ccc}
       0  &  4  & 9   \\   
       2  &  8  & 0   \\
       3  &  0  & 0
  \end{array}  
  \right) \ ; \
 A_5 \ .* \ W = \left(      
      \begin{array} {ccc}
       1  &  0  & 9   \\   
       2  &  8  & 0   \\
       0  &  12  & 0
  \end{array}  
  \right)  
\end{equation*} 

By computation, we evaluate that:
\begin{align}
&g_1(A_1 \ .* \ W) =  5   ||  10   ||  27    || 3  ||  12   || 27     \   \     &g_1(A_2 \ .* \ W)  =  5  || 20   || 12    ||  3   ||  16   ||  18        \notag   \\
&g_1(A_3 \ .* \ W)  =  5  || 26  ||  3   || 4   ||  12  ||  18        \    \     &g_1(A_4 \ .* \ W) =  13 ||  10   ||  3    || 5  ||  12   || 9               \notag   \\
&g_1(A_5 \ .* \ W)  =  10  || 10   || 12  ||  3   ||  20   ||  9    \    \      &                              \    \notag
\end{align}

It is easy to verify that  $A_1(2, 2)   \ne  A_2(2 , 2)$,  $A_1(2,  3)   \ne  A_2( 2 ,  3)$ ,  $A_1(3 ,  2)   \ne  A_2(3 , 2)$  and  $A_1(3 , 3)   \ne  A_2( 3 , 3)$.  All these differences imply that
 \begin{itemize}  
 \item  the second term of  $g_1(A_1 \ .* \ W) $ is not equal to the second term of $g_1(A_2 \ .* \ W)$,
 \item  the third term of  $g_1(A_1 \ .* \ W) $ is not equal to the third term of $g_1(A_2 \ .* \ W)$,
 \item  the fifth term of  $g_1(A_1 \ .* \ W) $ is not equal to the fifth term of $g_1(A_2 \ .* \ W)$,
 \item  the sixth term of  $g_1(A_1 \ .* \ W) $ is not equal to the sixth term of $g_1(A_2 \ .* \ W)$.
\end{itemize}  

More formally, from the construction of $g_1$, we can deduce easily that if $A(i,j)  \ne B(i,j)$, then :
\begin{description}
\item[$c^\circ$)]   the i-th term of  $g_1(A  \ .* \ W)$ would probably be different from the i-th term of $g_1(B \ .* \ W)$,
\item[$d^\circ$)]   the (n+j)-th term of  $g_1(A  \ .* \ W)$ would probably be different from the (n+j)-th term of $g_1(B \ .* \ W)$.
\end{description}

From the fact that $3DCT$ which is related to $g_2$ (this is an extension of $g_1$) is NP-Complete, we deduce that:
\begin{description}
\item[$e^\circ$)]   Given $y$ and a matrix  $W$, find a matrix $A$ such that  $g_2(A  \ .* \ W) = y$ is NP-Complete. 
\end{description}

Our idea is to build a new hash function $H_3$  such that $H_3 = H_2 o H_1$ where

\begin{itemize}  
 \item  $H_2$ is a classical hash function such as MD5, SHA-0, SHA-1, RIPEMD, HAVAL,...
 \item  $H_1$ is a function which exploits the ideas presented in $c^\circ$) , $d^\circ$) and $e^\circ$).
\end{itemize}  
                                                            
Let us denote $VOnes(n)$ the vector  such that $VOnes(n)  \  \in  \{ 0 , 1  \}^{n}$  and each of its elements is equal to 1. Also,  let us denote $MOnes(n)$ the matrix  such that $MOnes(n)  \  \in  \{ 0 , 1  \}^{n \times n}$ and each of its elements is equal to 1. in other words:
\begin{align}  \notag
&VOnes(n)_{i}    \    \     \    =  \  1 \    {\rm  where}   \ 1  \leq  i  \leq  n              \notag  \\  
&Mones(n)_{i,j}  \    =  \  1 \    {\rm  where}   \ 1  \leq  i , j    \leq  n        \notag
\end{align} 

We denote $N_+$ the set of strictly positive natural number  defined as follows:
\begin{equation}       \notag
N_+  =  \mathbb{N}     \setminus   \{ 0  \}   =    \{ 1, 2 , 3 , 4 ,  \dots      \}              \notag
\end{equation}
  
In the next sub-section, we formalize the observation made in points $c^\circ$) and $d^\circ$) and we take into account 
the NP-Completeness of 3DCT  to   build a new hash function.   

\subsection{Construction of the new hash function}

For any integers $a$ and $p$ such that $0 \leq a \leq -1+2^p$, let us denote $bin(a,p)$ the decomposition of the integer $a$ in base 2 on $p$ positions.
 In other words: 
\begin{equation}  \notag
bin(a,p) = x_{p-1} x_{p-2} \dots x_1 x_0 \ \ \ {\rm and} \ \ \ \sum_{i=0}^{p-1}  x_i \times 2^i = a
\end{equation}
Let us also define the following function:
\begin{equation}     
f_0(n)  =  \lceil log_2(n+1) \rceil     \notag
\end{equation}
$f_0(n)$  represents  the number of bits necessary to represent any integer between $0$ and $n$ in base 2.  \\  

We also define the following  functions:
\begin{align}
        f_1(A)   &=  max  \left\{\,     \sum_{j=1}^{n}  \sum_{k=1}^{n}    A(i, j, k )  \   :   \   1  \leq  i   \leq   n     \,\right\} ,       \notag    \\
        f_2(A)   &=  max  \left\{\,     \sum_{i=1}^{n}  \sum_{k=1}^{n}    A(i, j, k )  \   :   \   1  \leq  j   \leq   n     \,\right\}        \notag ,     \\
        f_3(A)   &=  max  \left\{\,     \sum_{i=1}^{n}  \sum_{j=1}^{n}    A(i, j, k )  \    :   \   1  \leq  k   \leq   n    \,\right\}   ,     \notag      \\
        f_4(A)   &=  max  \left\{\,     f_1(A)  ,  f_2(A)  ,  f_3(A)       \,\right\}.       \notag    
 \end{align}
$f_4(A)$ represents the maximun of sum of any $n$ consecutive elements of the matrix $A$ belonging to the same row, or to the same column or to the same file.  $f_0 o f_4(A)$ represents the number of bits necessary to represent in base 2 the sum of any $n$ consecutive elements of the matrix $A$ belonging to the same row, or to the same column, or to the same file.   \\

Subsequently, in the aim to be more precise, we redefine $g_2$ as follows:

\begin{align}
g_2(A) =  &bin ( R(1,1) , f_0 o f_4(A) )   ||  bin ( R(1,2) ,  f_0 o f_4 (A))    || \dots   ||  bin ( R(n,n) , f_0 o f_4(A) )   ||   \notag  \\
                  &bin ( C(1,1) , f_0 o f_4(A) )   ||  bin ( C(1,2)  , f_0 o f_4 (A))    || \dots   ||  bin ( C(n,n) , f_0 o f_4(A))    ||   \notag  \\       
                  &bin (  F(1,1) , f_o o f_4(A))    ||  bin ( F(1,2)  , f_0 o f_4(A))      || \dots   || bin (  F(n,n) , f_0 o f_4(A))         \notag 
\end{align}

Let us define the following problem:    \\

{\bf Problem 5:}     \\
$\phantom{salutsa}$ {\it Instance:}  A positive integer $n$,  two  binary strings $x$ and $y$    \\
$\phantom{salutsalutsalutsr l}$ two matrices  $V, W  \   \in  \mathbb{N}^{n \times  n \times n}$     \\
$\phantom{salutsa}$ {\it Query:}   Find  two matrices  $A, B \  \in  \{ 0 , 1 \}^{n \times  n \times n}$     \\
$\phantom{salutsaQuyqrg}$ \ such \  that: 
\begin{align}
       A                         &\ne      B                                         \notag      \\
       g_2(  A  \ .* \   V   )   &=   g_2 ( B  \ .* \    V    )  =   x        \notag      \\
       g_2(  A  \ .* \   W   )   &=   g_2 ( B  \ .* \   W   )   =  y        \notag        
 \end{align}
 
 Let us characterize the complexity of $Problem 5$.
 
\begin{proposition}     \label{ref:prop1}
 Problem 5  is  NP-Complete.
\end{proposition}
{\bf   Proof  Idea of Proposition  \ref{ref:prop1}:  }       \\
We want to show how to transform a solution of 3DCT to a solution of Problem 5
. Without loss of generality, we work in dimension 2. Let us suppose that we want to find a matrix
$A  \in  \{ 0 , 1 \}^{3 \times 3}$ such that:

\begin{subequations}   \label{E:gpgp}
\begin{gather}    
        \sum_{j= 1}^{3} A(i,j)   =    R(i)           \label{E:gp1}            \\
        \sum_{i= 1}^{3} A(i,j)   =    C(j)           \label{E:gp2}
 \end{gather}
 \end{subequations}
where $R = (3, 2, 1)$ and $C=(2, 3, 1)$.        \\
It is easy to see that the determination of the matrix  $A  \in  \{ 0 , 1 \}^{3 \times  3}  $ which verifies Equations (\ref{E:gpgp}) is also equivalent to determining the matrix $B  \in  \{ 0 , 1 \}^{6 \times  6} $ such that:  

\begin{subequations} \label{E:gh}
\begin{gather}     
    \sum_{j=1}^{6}  B(i,j)   =    Rd(i)     \label{E:gh1}       \\
    \sum_{i=1}^{6}  B(i,j)   =    Cd(j)     \label{E:gh2}
 \end{gather}
 \end{subequations}
where   $Rd = (3,  2,  1,  3, 2, 1 )$ and $Cd =(2,  3,  1,  2,  3,  1)$.      \\
{\bf   Remark  1:} $Rd$ ( respectively $Cd$ ) is a duplication of $R$ ( respectively  $C$  ).     \\
It is easy to see that  from the matrix: 

\begin{equation*}
 A =  \left(      
      \begin{array} {ccc}
       1    &    1    &   1   \\   
       1    &    1    &   0   \\
       0    &    1    &   0
  \end{array}  
  \right)  
\end{equation*} 
 which verifies Equations   (\ref{E:gpgp}), we can associate the two following matrices $B_2$ and $B_3$ 
\begin{equation*}
 B_2 =  \left(      
      \begin{array} {cc}
                  A                     &      0_ {3 \times  3 }  \\   
        0_ {3 \times  3 }     &              A    
  \end{array}  
  \right)  =  \left(      
      \begin{array} {cccccc}
       1  &  1  & 1  &  0  & 0 & 0   \\   
       1  &  1  & 0  &  0  & 0 & 0   \\
       0  &  1  & 0  &  0  & 0 & 0    \\   
       0  &  0 &  0  & 1   & 1 & 1    \\   
       0  &  0 &  0  & 1   & 1 & 0    \\   
       0  &  0 &  0  & 0   & 1 & 0     
  \end{array}  
  \right)  
\end{equation*}
 
\begin{equation*}
 B_3 =  \left(      
      \begin{array} {cc}
                   0_ {3 \times  3 }          &                  A                          \\   
                                 A            &      0_ {3 \times  3 }   
  \end{array}  
  \right)  =  \left(      
      \begin{array} {cccccc}
       0  &  0  &  0  &  1  &  1  & 1      \\   
       0  &  0  &  0  &  1  &  1  & 0     \\
       0  &  0  &  0  &  0  &  1  & 0      \\   
       1  &  1  &  1  &  0  &  0 &  0      \\   
       1  &  1  &  0  &  0  &  0 &  0      \\   
       0  &  1  &  0  &  0  &  0 &  0      
  \end{array}  
  \right)  
\end{equation*}
 which verify  Equations  (\ref{E:gh}). This is the idea of the transformation which associates to one solution of the problem defined in Equations (\ref{E:gpgp})  two  distinct 
 solutions of the problem defined in Equations (\ref{E:gh}).  \\    
 Before the  proof, let us introduce the function duplic (which is pseudo-duplication ) of x. We note:
 \begin{equation}
 x = x(1) x(2)  \dots  x(p)     
 \end{equation}
 where $x(i)  \in  \{ 0 , 1 \}$  and  $p = 3  \times n^2  \times  \lceil log_2(n+1) \rceil$. We define the function $t$ as follows:
 \begin{equation}
 t(i , n) =  i   \times n \times  \lceil log_2(n+1) \rceil  \ ; \ \ 0 \leq i \leq  3n.  
 \end{equation}  
The function duplic is defined as follows:
\begin{align}
duplic(x , n) =  &dcopy(x, 1 , n) || dcopy( x,2 , n) || \dots  || dcopy (x , n , n )  ||   \notag  \\
                 &dcopy(x, 1 , n) || dcopy( x,2 , n) || \dots  || dcopy (x , n , n )  ||   \notag  \\
                 &dcopy(x, n+1 , n ) || dcopy( x, n+2 , n ) || \dots  || dcopy (x , 2n , n )  ||   \notag  \\
                 &dcopy(x, n+1 , n ) || dcopy( x, n+2 , n) || \dots  || dcopy (x , 2n , n )  ||   \notag  \\
                 &dcopy(x, 2n+1 , n) || dcopy( x, 2n+2 , n) || \dots  || dcopy (x , 3n , n )  ||   \notag  \\
                 &dcopy(x, 2n+1 , n) || dcopy( x, 2n+2 , n) || \dots  || dcopy (x , 3n , n )     \notag
\end{align}                 
where $dcopy(x, i, n)$  is defined as follows:                     
\begin{eqnarray}   \notag
dcopy(x,i , n) =  strcopy(x, i, n ) || strcopy(x, i, n)
\end{eqnarray}
and
\begin{eqnarray}  \notag
strcopy(x, i, n) = x(1+t(i-1, n)) x(2+t(i-1, n)) \dots x(t(i,n))  
\end{eqnarray}  
For  illustration, $duplic(x, 3)$ is defined as follows:
\begin{align}
duplic(x,3)=&x(1) \dots x(6)x(1) \dots x(6)x(7) \dots x(12)x(7)\dots x(12)x(13) \dots x(18)x(13)\dots x(18) || \notag \\
&x(1) \dots x(6)x(1) \dots x(6) x(7) \dots x(12)x(7)\dots x(12)x(13) \dots x(18)x(13)\dots x(18) || \notag  \\          &x(19) \dots x(24) x(19) \dots x(24) x(25) \dots x(30) x(25) \dots x(30)x(31) \dots x(36)x(31) \dots x(36) || \notag \\
&x(19) \dots x(24) x(19) \dots x(24) x(25) \dots x(30) x(25) \dots x(30)x(31) \dots x(36)x(31) \dots x(36) ||  \notag \\
&x(37) \dots x(42) x(37) \dots x(42) x(43) \dots x(48) x(43) \dots x(48) x(49) \dots x(54) x(49) \dots x(54)|| \notag \\
&x(37) \dots x(42) x(37) \dots x(42) x(43) \dots x(48) x(43) \dots x(48) x(49) \dots x(54) x(49) \dots x(54) \notag   \end{align}

{\bf Remark 2:} In the definition of $strcopy(x,i,n)$, the term $x(1+t(i-1,n))x(2+t(i-1,n)) \dots x(t(i,n))$ means the concatenation of all the elements between $x(1+t(i-1,n))$ and $x(t(i,n))$. In other words: $x(1+t(i-1,n))x(2+t(i-1,n) \dots x(t(i,n))= x(1+t(i-1,n))x(2+t(i-1,n) \dots x(j) \dots x(-1+t(i,n))x(t(i,n))$ where $1+t(i-1,n) \leq j \leq t(i,n)$.

{\bf Proof  of  Proposition  \ref{ref:prop1} } : It suffices to show that $3DCT   \leq^P_m  Problem \  5$.   \\

Let us suppose that  the procedure  {\it Generalize}   solves   Problem 5  and we want to show how to build a procedure  {\it   Sol3DCT} which solves 3DCT.  \\
The procedure {\it  Sol3DCT} takes as input a binary string x, an integer n and returns as output the matrix $A$ of size $n$ such that $g_2(A) = x$. The procedure  {\it Generalize}
takes as input:   
\begin{itemize}  
 \item     p  the dimension of the matrices
 \item     two binary strings $x$ and  $y$  
 \item     two matrices $V$ and $W$
\end{itemize}  
and returns as output: 
\begin{itemize}  
 \item  two matrices $C$ and  $D$  such that:  
 \item  $g_2( C \ .* \ V ) =  g_2 ( D \ .* \ V ) =  x$ and $g_2( C \ .* \ W ) = g_2 ( D \ .* \ W ) = y$         
\end{itemize}  
We show in the  procedure below how to use $Generalize$  as a subroutine to solve $Sol3DCT$.
\begin{tabbing}      
Procedure \=  \textbf{ Sol3DCT(  n : integer  ,  x : string  ,   var   A  :  matrix   ) ;  } \\
\>   V,  W  , C  , D : matrix    \\
\>    p  ,  i  ,  j ,  k   :    integer        \\
\>   z       :   string         \\     
\> 	\textbf{begin} \=  \\
1: \> \>                $V      \longleftarrow        MOnes(2n) $                    \\
2: \> \>		$W      \longleftarrow        Mones(2n)$       \\
3: \> \>		$z      \longleftarrow          duplic(x, n)$       \\
4: \> \>		$p       \longleftarrow          2  \times   n$       \\
5: \> \>		$Generalize ( p , z , z, V , W ,  C , D   )$      \\ 
6: \> \>  For i = 1 \= to  n   do       \\
7:\> \> \> For j = 1 \=  to  n   do          \\
8:\> \> \>\>   For k = 1 \=  to  n   do                 \\
9:\> \> \>\> \>  $A(i, j, k )   \longleftarrow  C(i,j,k)  +  C(i, j+n, k)$         \\
10:\> \> \>\>   Endfor              \\
11:\> \> \> Endfor          \\
12: \> \>  Endfor       \\  
 \>  \textbf{end}   \\
\end{tabbing}
{\bf   Remark 3:}   In the procedure  \textbf{ Sol3DCT}, the matrix  $A$  belongs to the set  $\{ 0 , 1 \}^{n \times  n \times n}$, whereas  the matrices  $C , D$  belong to the  set  $\{ 0 , 1 \}^{2n \times  2n \times 2n}$.   \\

The string $z$  of the procedure   \textbf{ Sol3DCT } (see instruction 3) is  constructed such that $g_2(A) = x$  if and only if  the matrices $C$  and $D$ defined in Equations (\ref{EE:Ep1})  and (\ref{EE:Ep2})  are the solutions of Problem  5 with the following entries: 
\begin{itemize}  
 \item     2n  the dimension of  the matrices,
 \item     two binary strings $z$ and  $z$, 
 \item     two matrices $V$ and $W$ such that  $V = Mones(2n)$ , $W = Mones(2n)$.
\end{itemize}  

The  terms  of  the  matrix  $C$  are:
\begin{equation}         \label{EE:Ep1}
      \begin{cases}
 C(i ,  j ,  k )  =   A( i , j , k ),              &\text{if  \  $1 \ \leq  \  i , j , k   \  \leq  \  n$ ; }      \\   
 C(i ,  j+n ,  k )  =   0,                        &\text{if  \  $1 \ \leq  \  i , j , k   \  \leq  \  n$  ; }       \\        
 C(i+n ,  j ,  k )  =   0,                        &\text{if  \  $1 \ \leq  \  i , j , k   \  \leq  \  n$ ;  }      \\             
 C(i+n ,  j+n ,  k )  =   A( i , j , k ),    &\text{if  \  $1 \ \leq  \  i , j , k   \  \leq  \  n$  ;  }      \\         
C(i ,  j ,  k+n )  =    0,                        &\text{if  \  $1 \ \leq  \  i , j , k   \  \leq  \  n$  ;  }      \\   
C(i ,  j+n ,  k+n )  =  A(i,j,k),            &\text{if  \  $1 \ \leq  \  i , j , k   \  \leq  \  n$  ;  }      \\        
 C(i+n ,  j ,  k+n)  = A(i,j,k),                 &\text{if  \  $1 \ \leq  \  i , j , k   \  \leq  \  n$  ;  }      \\             
 C(i+n ,  j+n ,  k+n)  =  0,                    &\text{if  \  $1 \ \leq  \  i , j , k   \  \leq  \  n$.}       
        \end{cases}
\end{equation}  

The  terms  of  the  matrix  $D$  are:
\begin{equation}         \label{EE:Ep2}
      \begin{cases}
D(i ,  j ,  k )  =   0,                             &\text{if  \  $1 \ \leq  \  i , j , k   \  \leq  \  n$ ;   }      \\   
D(i ,  j+n ,  k )  =   A(i, j, k),             &\text{if  \  $1 \ \leq  \  i , j , k   \  \leq  \  n$ ;    }      \\        
D(i+n ,  j ,  k )  =   A(i, j, k),             &\text{if  \  $1 \ \leq  \  i , j , k   \  \leq  \  n$ ;    }      \\             
D(i+n ,  j+n ,  k )  =   0,                   &\text{if  \  $1 \ \leq  \  i , j , k   \  \leq  \  n$ ;    }      \\         
D(i ,  j ,  k+n )  =  A(i, j, k),              &\text{if  \  $1 \ \leq  \  i , j , k   \  \leq  \  n$ ;    }     \\   
D(i ,  j+n ,  k+n )  =  0,                    &\text{if  \  $1 \ \leq  \  i , j , k   \  \leq  \  n$ ;    }      \\        
D(i+n ,  j ,  k+n )  =  0,                    &\text{if  \  $1 \ \leq  \  i , j , k   \  \leq  \  n$ ;    }      \\             
D(i+n ,  j+n ,  k+n )  =  A(i, j, k),     &\text{if  \  $1 \ \leq  \  i , j , k   \  \leq  \  n$.   }       
        \end{cases}
\end{equation}  
$\blacksquare$  \\

The main idea of the design of  the  Collision-resistant hash function $H_3$ is that: 
\begin{itemize}  
\item the hash function $H_3$ is the composition of two functions $H_1$ and $H_2$,
\item the function $H_1$ is a function for which {\it Problem 1}, {\it Problem 2} and {\it Problem 3} can't be solved efficiently and $H_1$ is not a compression function. 
\item $H_2$ is a hash function such as SHA-256, RIPEMD, or HAVAL, ....   
\end{itemize}
  \begin{notation}  
 Let us consider two vectors $V_1$ and $V_2$.  We say that $V_1$ is not a linear combination of $V_2$  and we note $V_1$  is  NLC  of  $V_2$  if and only if 
   $\not\exists  \  \alpha \  \in \mathbb{R}$ such that   $V_1  =  \alpha V_2$. 
  \end{notation}  

Two  matrices  $F, G  \in  N_+^{n \times  n \times n}$ verify the hypotheses (\ref{E:gp})  if and only if :
\begin{subequations}     \label{E:gp}  
\begin{gather}
  F    \ne   \    G                                        \label{E:gp1}                                                                                                         \\  
  \forall  i , j   {\rm   \  such  \   that  \   }   \    1  \leq  i, j  \leq  n , {\rm     \  the  \  vector   \   F(i, j, *)   \   is  \  NLC  \   of  \  the  \  vector   \  G(i, j, *) }        \label{E:gp2}               \\ 
  \forall  j , k  {\rm   \  such   \  that   \   }   \    1  \leq  j, k \leq  n ,  {\rm  \    the  \   vector  \  F(*, j, k)  \  is  \  NLC \  of \  the \  vector  \  G(*, j, k)  }          \label{E:gp3}               \\ 
  \forall  i , k  {\rm   \  such  \  that  \   }   1   \   \leq  i, k \leq  n , {\rm \  the  \  vector  \  F(i,  *, k)  \  is \  NLC \  of \   the \  vector \  G(i,  *, k)  }        \label{E:gp4}               \\       
  \forall  i , j   {\rm   \  such  \  that   \   }   \   1  \leq  i, j  \leq  n , {\rm \   the  \  vector  \  F(i, j, *) \   is \  NLC \  of  \  the \  vector \  VOnes(n)  }          \label{E:gp5}               \\ 
  \forall  j , k  {\rm   \  such  \  that  \   }   \   1  \leq  j, k \leq  n , {\rm   \  the \   vector  \  F(*, j, k)   \   is \   NLC   \  of   \  the  \ vector  \ VOnes(n)  }          \label{E:gp6}               \\ 
  \forall  i , k  {\rm   \  such  \  that  \   }   \   1  \leq  i, k \leq  n , {\rm   \  the \   vector   \  F(i,  *, k)  \  is \    NLC   \  of   \  the  \  vector   \  VOnes(n)   }       \label{E:gp7}        \\    
 \forall  i , j   {\rm   \  such  \  that   \   }   \   1  \leq  i, j  \leq  n , {\rm \   the  \  vector  \  G(i, j, *) \   is \  NLC \  of  \  the \  vector \  VOnes(n)  }          \label{E:gp8}               \\ 
 \forall  j , k  {\rm   \  such  \  that  \   }   \   1  \leq  j, k \leq  n , {\rm   \  the \   vector  \  G(*, j, k)   \   is \   NLC   \  of   \  the  \ vector  \ VOnes(n)  }          \label{E:gp9}               \\ 
 \forall  i , k  {\rm   \  such  \  that  \   }   \   1  \leq  i, k \leq  n , {\rm   \  the \   vector   \  G(i,  *, k)  \  is \    NLC   \  of   \  the  \  vector   \  VOnes(n)   }       \label{E:gp10}                     
\end{gather}
\end{subequations}

The matrices $V$ and $W$ used  as entries in the procedures $H_1$ and $H_3$ below verify the hypotheses defined by Equations  (\ref{E:gp}). We note $\epsilon$ the empty chain.  
Let us define the function $VectMat$ which takes as input a vector $Vect$ of size $n^3$ and returns as output an equivalent matrix $A$ of size $n \times n \times n$. 


{\bf  Procedure} $VectMat$ (Vect : Table[1..$n^3$] of bit ; Var A : Table[1..n, 1..n, 1..n] of bit )   \\
$\phantom{salut}$ Var  i, j, k , t  : integer  \\
$\phantom{salut}$ Begin          \\
$\phantom{salutsal}$$t  \leftarrow  1$   \\
$\phantom{salutsal}$For i = 1 to  n  do   \\
$\phantom{salutsalutsa}$For j = 1 to  n  do \\
$\phantom{salutsalutsalut}$For k = 1 to n  do   \\
$\phantom{salutsalutsalutsalut}$$A(i,j,k)  \leftarrow  Vect(t)$    \\
$\phantom{salutsalutsalutsalut}$$t  \leftarrow  t+1$    \\
$\phantom{salutsalutsalut}$endfor   \\
$\phantom{salutsalutsa}$endfor       \\
$\phantom{salutsal}$endfor         \\
$\phantom{Salut}$End              \\

The function $H_1$ is defined as follows:   \\

{\bf Function } $H_1$ :   \\
$\phantom{salut}$ {\it Entry.} $M_0$ the initial message  \\
$\phantom{salutsalutius}$ $V$ : Table[1..n, 1..n,1..n] of integer    \\
$\phantom{salutsalutius}$ $W$ : Table[1..n, 1..n,1..n] of integer    \\
$\phantom{salutsaluts}$ $n$: an integer   \\
$\phantom{salut}$ {\it Output.} $M_2$: an intermediate message    \\
$\phantom{salut}$ Var  i, p   : integer  \\
$\phantom{salut}$ Begin     \\
$\phantom{salutsalu}$ 1. Pad $M_0$ with one bit equal to 1, followed by a variable number of \\
$\phantom{salutsadasut}$ zero bits and a block of bits encoding the length of $M_0$ in bits, \\
$\phantom{salutsadasut}$ so that the total length of the padded message is the smallest   \\
$\phantom{salutsaADh}$ possible multiple of $n^3$. Let $M_1$ denote the padded message  \\
$\phantom{salutsalu}$ 2. Cut $M_1$ into a sequence of $n^3$-bits vectors \\
$\phantom{salSDutsKJLalu}$ $B_1, B_2, \dots, B_i, \dots B_{p}$  \\
$\phantom{salutsalu}$ 3. $M_2 \ \leftarrow$ \ $\epsilon$      \\
$\phantom{salutsartul}$4.  For i = 1 to p do   \\
$\phantom{salutsfagjalutsa}$ 4.1 $VectMat(B_i, A)$ \\
$\phantom{salutsalhggsutsa}$ 4.2 $M_2$  $\leftarrow$ $M_2$ $\|$ $g_2( A  .*  V )$ $\|$ $g_2(  A  .*  W  )$   \\
$\phantom{salutgsgglkggu}$ Endfor \\
$\phantom{salutsalu}$ 5. return $M_2$ \\
$\phantom{Saleut}$End              \\

Our hash function $H_3$ is defined as the composition of the function $H_1$ and $H_2$, where $H_2$ is a hash function
 such as SHA-256, RIPEMD, HAVAL...   The matrices  $V$ and $W$ used as entry  in the  hash function  $H_3$  must verify the hypotheses defined in Equations (\ref{E:gp}).    
To obtain the hash of the message $M_0$ by $H_3$, we proceed as follows:
\begin{itemize}
\item we obtain the intermediate message $M_2$ by application of the function $H_1$ to the message $M_0$,
\item by application of the hash function $H_2$ to $M_2$, we build the hash of the initial message. 
\end{itemize}
Formally, the hash function is defined as follows: \\ \\
{\bf Procedure} $H_3$ :   \\
$\phantom{salut}$ {\it Entry.} $M_0$  the initial message   \\
$\phantom{salutsalutius}$ $V$ : Table[1..n, 1..n,1..n] of  integer    \\
$\phantom{salutsalutius}$ $W$ : Table[1..n, 1..n,1..n] of  integer    \\
$\phantom{salutsalutos}$ $n$ : an integer   \\
$\phantom{salut}$  {\it Output.} $Result$ : the hash of the message $M_0$    \\
$\phantom{salut}$ Begin     \\
$\phantom{salutsalutsal}$ $M_2 \ \leftarrow \ H_1 (M_0 ,  V ,  W ,  n)$   \\
$\phantom{salutsalutsal}$ $Result \ \leftarrow \ H_2 (M_2) $   \\
$\phantom{SalutS}$End              \\

{\bf Comment  :}  \\

We can represent roughly the function $H_1$ as follows:
\newpage
\begin{figure}[htbp]
\centering
\epsfxsize=8.6cm
\epsfbox{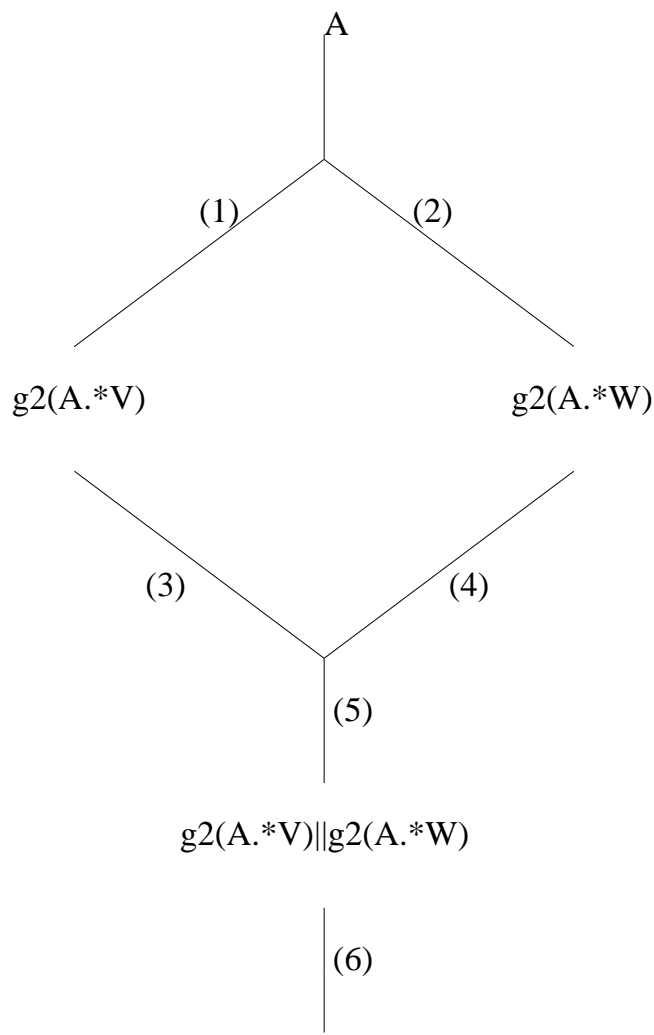}
\caption{\label{figgraph9} {\em Roughly Representation of the function $H_1$  }}
\end{figure}
In the Figure \ref{figgraph9}: 
\begin{itemize}
\item the aim of the branches (1) and (2) is to make that the Problem 2 and Problem 3 are difficult to solve efficiently for the function $H_3$ 
\item the aim of the branch (6) is to make sure that Problem 1 is difficult to solve efficiently for the function $H_3$
\end{itemize}

During some  attacks, an adversary is needed to solve the following problem:

{\bf Problem 6:}     \\
$\phantom{salutsa}$ {\it Instance:}  Matrices  A, V,  W      \\
$\phantom{salutsalutsalutsr salut}$ Binary strings : $g_2(  A  \ .* \   V   )$   and  $g_2(  A  \ .* \   W   )$      \\
$\phantom{salutsa}$ {\it Query:}   Find  a matrix  $B$ such that  $A  \ne  B$  and: 
\begin{align}
       g_2(  A  \ .* \   V   )  &=   g_2 ( B  \ .* \    V    )         \notag      \\
       g_2(  A  \ .* \   W   )  &=   g_2 ( B  \ .* \   W   )       \notag        
 \end{align}
Based on Problem 5, we deduce that $Problem  \  6$ is NP-Complete.   \\

 Second Preimage  attack and Collision of the function $H_3$ are difficult because:
\begin{itemize}      
 \item   Problem 5  and Problem 6  are NP-Complete,    
  \item  From the fact that $V$ and $W$ verify the hypotheses (\ref{E:gp}), we deduce that  if we take two matrices $A$ and $B$ such that  $A  \  \ne  \  B$, then we would  probably have $g_2(  A  \ .* \   V )  ||  g_2(  A  \ .* \  W   )  \ne   g_2 ( B  \ .* \  V    )    ||   g_2 ( B   \ .* \   W   ).$
\end{itemize} 

First Preimage attack of the function $H_3$ is difficult because  the  3DCT is NP-Complete. Truncated  differential attack of $H_1$ is possible, but  the differential attack of $H_3$ is difficult
 because  3DCT is NP-Complete and also Problem 5 is  NP-Complete.
 
 \section{Numerical  Simulation}
 
Let's consider the two messages x1 and x2 :\\
\begin{align}
x1 = &d1 31 dd 02 c5 e6 ee c4 69 3d 9a 06 98 af f9 5c    \notag   \\
     &2f ca b5 87 12 46 7e ab 40 04 58 3e b8 fb 7f 89    \notag   \\
     &55 ad 34 06 09 f4 b3 02 83 e4 88 83 25 71 41 5a    \notag   \\
     &08 51 25 e8 f7 cd c9 9f d9 1d bd f2 80 37 3c 5b    \notag   \\
     &d8 82 3e 31 56 34 8f 5b ae 6d ac d4 36 c9 19 c6    \notag   \\
     &dd 53 e2 b4 87 da 03 fd 02 39 63 06 d2 48 cd a0    \notag   \\
     &e9 9f 33 42 0f 57 7e e8 ce 54 b6 70 80 a8 0d 1e    \notag   \\
     &c6 98 21 bc b6 a8 83 93 96 f9 65 2b 6f f7 2a 70    \notag
\end{align}

\begin{align}
x2 = &d1 31 dd 02 c5 e6 ee c4 69 3d 9a 06 98 af f9 5c  \notag  \\
     &2f ca b5 07 12 46 7e ab 40 04 58 3e b8 fb 7f 89  \notag  \\
     &55 ad 34 06 09 f4 b3 02 83 e4 88 83 25 f1 41 5a  \notag  \\
     &08 51 25 e8 f7 cd c9 9f d9 1d bd 72 80 37 3c 5b  \notag  \\
     &d8 82 3e 31 56 34 8f 5b ae 6d ac d4 36 c9 19 c6  \notag  \\
     &dd 53 e2 34 87 da 03 fd 02 39 63 06 d2 48 cd a0  \notag  \\
     &e9 9f 33 42 0f 57 7e e8 ce 54 b6 70 80 28 0d 1e  \notag  \\
     &c6 98 21 bc b6 a8 83 93 96 f9 65 ab 6f f7 2a 70  \notag
\end{align}

We have MD5(x1)=MD5(x2)= EFE502F744768114B58C8523184841F3 \\
after applying our hash function on these messages using $n=8$, $V[i][j][k] = i + 8j + 64k$, $W[i][j][k] = 700-(j+8*k+64*i)$ for $1\leq i\leq n$ we obtain:\\
$H_3$(x1)=  5fe0e56f9a4ab66a47d73ce660a2c4eb and \\
$H_3$(x2) = 620e2f3cfe0afc403c0a8343173526fc. \\
It follows that $MD5(x1)=MD5(x2)$ whereas $H_3(x1) \ne H_3(x2)$.

\section{Conclusion} 
From a classical hash function $H_2$, we have built a new hash function $H_3$ from which First Preimage attack, Second Preimage attack and Collision attack are difficult to solve.  Our new hash 
function is a composition of functions. The construction used the NP-completeness  of Three-dimensional contingency tables and the relaxation of the  constraint that a 
 hash function should also be a compression function. The complexity of our new hash function increases with regard to the complexity of classical hash functions.

\bibliographystyle{fplain}

\begin{thebibliography}{2}
\bibitem{Bru 80} R. A. Brualdi, {\it Matrices of Zeros and Ones with Fixed Row and Column Sum Vectors}, Linear Algebra and its Applications,
 {\bf 33}, 1980, pp. 159-231.
\bibitem{Cox 80} L. Cox, {\it Suppression methodology and statistical disclosure control}, J. Amer. Statist. Assoc., 75(1980), pp. 377-385.
\bibitem{Fel 72} I. P. Fellegi, {\it On the question of statistical confidentiallity}, J. Amer. Statist. Assoc., 67, (1972), pp. 7-18.
\bibitem{Ful 60} D. R. Fulkerson, {\it An upper bound for the permanent of a fully indecomposable matrix}, Pacific J. Math.,
  {\bf 10}, 1960 , pp. 831-836.
\bibitem{Gal 57} D. Gale, {\it A Theorem on flows in networks}, Pacific J. Math., {\bf 7}, 1957 , pp. 1073-1082.
\bibitem{Hab 63} R. M. Haber, {\it Minimal term rank of a class of (0,1)-matrices},  Canad. J. Math., {\bf 15}, 1963 ,
 pp. 188-192.
\bibitem{HYXW 07} Hongbo Yu, Xiaoyun Wang, {\it Multi-Collision Attack on the Compression Functions of MD4 and 3-Pass Haval},
 Lecture Notes in Computer Science, {\bf 4817} , Springer 2007, pp. 206-226.
\bibitem{HY 05} Hongbo Yu, Gaoli Wang, Guoyan Zhang, Xiaoyun Wang, {\it The Second-Preimage Attack on MD4},
 Lecture Notes in Computer Science, {\bf 3810} , Springer 2005, pp. 1-12.
\bibitem{Kao 93} M. -Y. Kao, D. Gusfield Hongbo Yu, {\it Efficient detection and protection of information in cross tabulated tables:
 Linear invariant set}, SIAM J. Disc. Math., 6 (1993), pp. 460-473.
\bibitem{XHY 05} Xiaoyun Wang,  Hongbo Yu, Yiqun Lisa Yin, {\it Efficient Collision Search Attack on SHA-0},
 Lecture Notes in Computer Science, {\bf 3621} , Springer 2005, pp. 1-16.
\bibitem{XW 05}  Xiaoyun Wang, Yiqun Lisa Yin, Hongbo Yu, {\it Finding Collisions in the Full SHA-1},
 Lecture Notes in Computer Science, {\bf 3621}, Springer 2005, pp. 17-36.
\bibitem{XXW 05} Xiaoyun Wang, Xuejia Lai, Dengguo Feng, Hui Cheng, Xiuyuan Yu, {\it Cryptanalysis of the Hash Functions
 MD4 and Ripemd}, Lecture Notes in Computer Science, {\bf 3494} , Springer 2005, pp. 1-18.
\bibitem{XWW 05} Xiaoyun Wang,  Hongbo Yu, {\it How to Break MD5 and Other Hash Functions},
 Lecture Notes in Computer Science, {\bf 3494} , Springer 2005, pp. 19-35.
\bibitem{BA 94} Bert den Boer,  Antoon Bosselaers, {\it Collisions for the compression functions of MD5},
 Lecture Notes in Computer Science, {\bf 765}, Springer 1994, pp. 293-304.
\bibitem{IJ 94} R. W. Irving, M. R. Jerrum, {\it Three-Dimensional Statistical Data Security Problems},
 SIAM J. Comput., Vol. {\bf 23}, No 1, pp. 170-184, 1994.
\bibitem{Rys 57} H. J. Ryser, {\it Combinatorial properties of matrices of zeros and ones},
 Canad. J. Math., Vol. {\bf 9}, pp. 371-377, 1957.
\bibitem{San 84} G. Sande, {\it Automated cell suppression to preserve confidentiality of business statistics},
                 Statist. J. United Nations ECE 2 (1984) pp. 33-41.
\end{thebibliography}

\end{document}